# Efficient method for controlling the spatial coherence of a laser

M. Nixon[1], B. Redding[2], A. A. Friesem[1], H. Cao[2] and N. Davidson[1,*]

[1]Dept. of Phys. of Complex Systems, Weizmann Institute of Science, Rehovot 76100, Israel
[2]Dept. of Appl. Phys., Yale University, New Haven, Connecticut 06520, USA
*Corresponding author: Nir.Davidson@weizmann.ac.il

An efficient method to tune the spatial coherence of a degenerate laser over a broad range with minimum variation in the total output power is presented. It is based on varying the diameter of a spatial filter inside the laser cavity. The number of lasing modes supported by the degenerate laser can be controlled from 1 to 320,000, with less than a 50% change in the total output power. We show that a degenerate laser designed for low spatial coherence can be used as an illumination source for speckle-free microscopy that is 9 orders of magnitude brighter than conventional thermal light.

One of the important characteristics of lasers is their spatial coherence. With high spatial coherence the light that emerges from the lasers can be focused to a diffraction limited spot or propagate over long distances with minimal divergence. Unfortunately, high spatial coherence also results in deleterious effects such as speckles, so the lasers cannot be exploited in many full-field imaging applications[1].

Traditional light sources usually operate at the two extremes, with lasers and super luminescent diodes exhibiting very high spatial coherence and thermal sources or light emitting diodes (LEDs) having very low spatial coherence. Yet, for many applications, a light source with partial or even tunable spatial coherence is needed. Often, sources with an intermediate degree of spatial coherence can be exploited before significant speckles are formed, depending on the scattering properties of the sample and the illumination and collection optics[2-4]. Such sources could provide some of the advantages of laser sources, such as improved photon degeneracy, directionality, spectral control, and efficiency, as compared to thermal sources or LEDs with low spatial coherence. Sources with partial spatial coherence are also able to mitigate noise due to coherent artifacts in digital holographic microscopes[5,6]. Recently, new imaging modalities, such as HiLo microscopy, combine information collected with both high and low spatial coherence sources to obtain images that are superior to those obtained with either source alone[7]. Thus, there is a considerable need for light sources offering intermediate spatial coherence and, in particular, sources in which the spatial coherence can be tailored to the application.

While the spatial coherence of LEDs or thermal sources can be increased through spatial filtering, their corresponding output energies are significantly reduced. A more efficient approach could be to reduce the spatial coherence of laser sources by placing a time varying optical diffuser in the output light path. Unfortunately, this involves moving parts and long acquisition times [8-10]. Recently, random lasers were shown to exhibit tunable spatial coherence[11] and the ones engineered for low spatial coherence were incorporated in full-field imaging applications to obtain images whose quality is similar to those obtained with traditional low spatial coherence sources such as LEDs[12]. Unfortunately, random lasers have relatively high lasing thresholds and poor collection efficiencies.

Here, we propose and demonstrate a novel and efficient approach for tuning the spatial coherence of a laser source. This tuning is achieved with a degenerate laser where the number of transverse modes supported by the laser can be controlled from one to as many as 320,000. Moreover, the output energy remains relatively constant over the entire tuning range of spatial coherence.

The degenerate laser[13] that can support many traverse lasing modes[14,15] is shown schematically in Fig. 1. It is comprised of a Nd:YAG gain medium lasing at 1064 nm, a flat back mirror with 90% reflectivity and a flat front mirror with 40% reflectivity. To reduce the effects of thermal lensing, pumping was achieved using two 100 µs pulsed Xenon flash lamps operating at 1Hz to obtain quasi CW operation (sub 100 µs pulses). Between the mirrors two lenses in a 4$f$ telescope configuration were inserted in order to image the front mirror plane onto the back mirror plane. This imaging configuration satisfies the degeneracy condition because the transverse electric field distribution is imaged onto itself after propagating one full round trip. Consequently, any transverse electric field distribution represents an eigenmode of the degenerate cavity. Since these eigenmodes have the same quality factor and path length, lasing occurs simultaneously in many traverse modes. We therefore expect that the number of transverse lasing modes will be proportional to the ratio of the gain medium area over the diffraction limited area, which is determined by the smallest NA of the elements inside the laser cavity. In our laser, the 10 mm diameter 10 cm long gain rod determines the limiting NA as 0.05, resulting in a diffraction limited spot size of ~10 µm on the mirror. Thus, our degenerate laser could, in principle, support ~$10^6$ transverse lasing modes, i.e. light with very low spatial coherence.

The direct relation between the NA and the number of transverse modes provides a natural method to control the spatial coherence. Specifically, by reducing the NA, fewer transverse modes will be supported until eventually only a single mode with high spatial coherence should be possible. One possibility for reducing the NA would be to effectively reduce the diameter of the gain medium, but this would also result in a corresponding reduction of output energy. Alternatively, the NA can be controlled by varying the diameter of a pinhole positioned at the focal plane in between the two lenses, directly inside the laser cavity. By decreasing the diameter of the pinhole, it is possible to reduce the effective NA and filter out the higher spatial frequencies *without* reducing the effective gain area. Accordingly, decreasing the diameter of the pinhole will significantly reduce the number of modes while maintaining the same output energy.

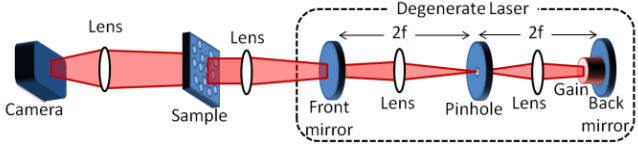

Fig. 1. (Color online) Experimental arrangement for controlling the number of lasing modes, and thereby the spatial coherence, of a degenerate laser. The degenerate laser consists of a gain medium, front and back flat mirrors, and two lenses in a 4f telescope configuration. The number of lasing modes is controlled by means of a variable pinhole aperture positioned at the focal plane between the lenses. In the imaging experiments, the collimated laser output illuminated a sample whose far-field speckle pattern was detected by means of a CCD camera.

We investigated our ability to tune the spatial coherence of our degenerate laser by adjusting the pinhole diameter while monitoring the total output energy and the ability to suppress speckles. Specifically, as shown in Fig. 1, the collimated light from the degenerate laser illuminated an optical diffuser and the far-field speckle pattern was detected by means of a charge coupled device (CCD) camera. The diffuser was designed with a random phase correlation length of 5 µm and no zero order transmittance (Newport 10° light shaping diffuser). To quantitatively compare speckle patterns, we calculated their speckle contrast $C$ as

$$C = \sigma_I / \langle I \rangle, \quad (1)$$

where $\sigma_I$ is the standard deviation and $\langle I \rangle$ is the average intensity. The results are presented in Fig. 2. Figure 2A shows the measured speckle contrast as a function of the pinhole diameter (solid red curve). As the pinhole diameter increases, more lasing modes are supported, providing lower spatial coherence and lower speckle contrast. By controlling the pinhole diameter, we were able to vary the speckle contrast from $C > 0.8$ for a 0.12 mm pinhole diameter to $C = 0.01$ for a 6 mm pinhole diameter.

We determined the number of lasing modes by measuring the beam quality $M^2$ of the laser output. The beam quality is given by the product of the beam width and the beam divergence angle relative to that of a diffraction limited single mode beam given by $\lambda/\pi$. In our system, we took the total number of modes $N$ to be $M^4$ (the product of $M^2$ along the *x*-axis and $M^2$ along the *y*-axis)[16]. Using a 0.12 mm pinhole diameter, we found that $M^4 = 1.06$, indicating lasing in a single $TM_{00}$ transverse mode. The measured speckle contrast for the 0.12 mm pinhole diameter was $C > 0.8$. This contrast of 0.8, which is lower than the expected contrast of $C = 1$, was comparable to that obtained with a separate single mode fiber laser (also operating at 1064 nm), indicating that the contrast was limited by the imaging setup, e.g. finite pixel sizes of the camera that average the speckle and reduce the contrast. As the pinhole diameter was increased, the $M^4$ and correspondingly the number of transverse laser modes also increased. A pinhole diameter of 6 mm resulted in multi-mode lasing, with a measured $M^4 > 320,000$ and a speckle contrast of $C = 0.01$.

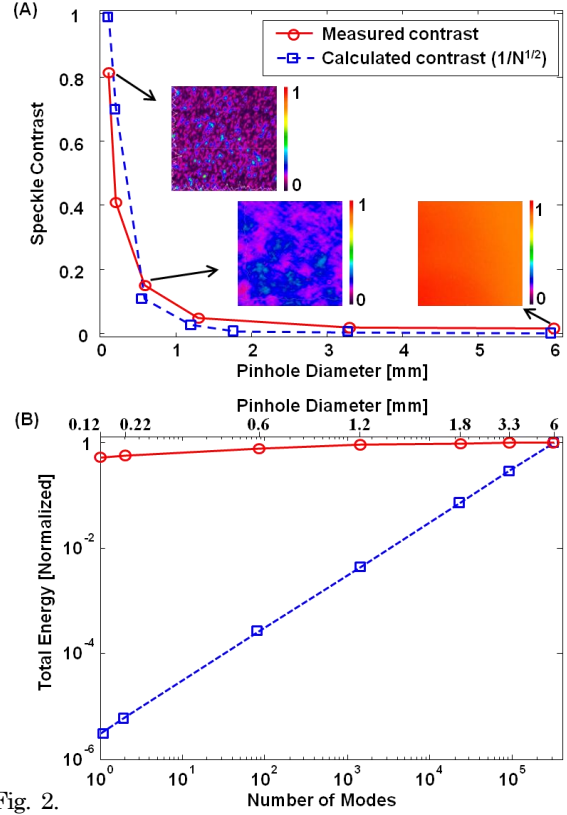

Fig. 2. (Color online) Experimental and calculated speckle contrast and laser output energy as a function of the pinhole diameter. (A) Experimental speckle contrast (red circles with solid line) and calculated speckle contrast (blue squares with dotted line) as a function of the pinhole diameter. (B) The measured output energy (red circles with solid line) as a function of the pinhole diameter and the corresponding number of modes, found by measuring the relative beam quality $M^4$. The blue squares with dotted line indicate the expected output energy if the number of modes would have been selected by means of spatial filtering outside the cavity, assuming the energy was equally distributed among all 320,000 lasing modes supported with the largest pinhole.

Using the measured $M^4$ values, we calculated the expected speckle contrast as $C = N^{-1/2}$ [1]. As evident in Fig. 2A, the calculated results are in good agreement with the experimentally measured contrast values. There are small differences due to the experimental limitations such as the finite resolution of imaging system, non-uniform illumination and camera noise which limited the minimum and maximum speckle contrast that can be measured.

We also investigated the effect of the pinhole diameter on the total output energy of the degenerate laser. The results are presented in Fig. 2(B). The output energy was normalized to the maximum value of 37 mJ. As evident, the total measured output energy varies only slightly as a function of the number of modes (solid red plot). For comparison, we added the dashed blue line denoting the expected total energy if we were able to select a subset of the modes by means of spatial filtering outside the cavity (assuming that the energy was evenly distributed among all 320,000 modes supported with the largest pinhole). As evident, the process of mode selection by means of the intra-cavity pinhole inside the degenerate laser is exceptionally efficient, because of the laser's ability to redistribute the energy into the remaining non-filtered modes with very little losses.

The remarkable efficiency of redistributing energy over 5 orders of magnitude of modes is a unique property of the degenerate cavity that cannot be obtained with conventional stable cavity resonators. As we will discuss in the next paragraph, for the degenerate cavity, all modes are degenerate so there is no intrinsic characteristic mode size, allowing the laser to adopt any mode size dictated by the pinhole. For stable cavity resonators, however, the characteristic mode size is dictated by their geometry, e.g. the curvature of a convex mirror. As a consequence, spatial filtering inevitably introduces loss in all modes due to the inherent mismatch between the cavity mode size and the spatial filter size.

The degenerate laser with tunable spatial coherence could be very advantageous as a high intensity illumination source for speckle-free, full-field imaging applications. To demonstrate this we performed experiments where we placed a U.S. Air Force resolution chart after the random diffuser and recorded its image with the camera. The results are presented in Fig. 3. Using a 120 µm diameter pinhole, the laser was restricted to single mode operation with high spatial coherence so the detected image was corrupted by high contrast speckles, as shown in Fig. 3A. However, when using a 6 mm diameter pinhole, the laser supported multimode lasing operation with low spatial coherence so the detected image was of high quality without coherent artifacts, as shown in Fig. 3B.

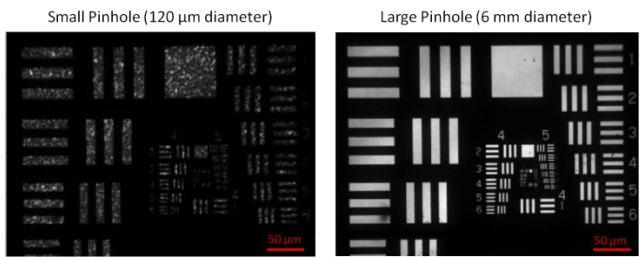

Fig. 3. Experimental demonstration for speckle-free full-field imaging with our degenerate laser. (A) Detected image of a resolution chart and diffuser, using a small pinhole diameter (single mode lasing with high spatial coherence). (B) Detected image using a large pinhole diameter (multi-mode lasing with low spatial coherence).

While LEDs or thermal sources could produce speckle-free images similar to those obtained using the 6 mm diameter pinhole, the degenerate laser is much brighter, enabling short integration times or imaging through lossy media.

We estimate that the photon degeneracy (a metric describing source brightness proportional to spectral radiance) of our laser when operating with >320,000 modes to be 9 orders of magnitude higher than typical LEDs or thermal sources[12] for long integration times (and 13 orders of magnitude brighter within the 100 µs duration lasing time).

To understand how our degenerate laser can have a broadly tunable spatial coherence, we calculated the electric field distribution in our laser cavity and its dependence on the pinhole diameter. Specifically, we resorted to a Fox-Li type algorithm[17] where an initial state (usually a random matrix) is propagated through the cavity iteratively, mimicking the effect of the circulating light, until a steady state eventually emerges. This steady state corresponds to the lowest threshold lasing mode in the cavity which, in general, depends on the initial conditions if more than one state exists. The propagation process involved manipulating a two-dimensional field matrix $E$ by appropriate operators that represent the various components that comprise the cavity, such as the gain aperture and pinhole. We accounted for gain saturation by using a gain matrix, $G$, defined as $G=p/(1+I/I_s)$, where $p$ is the pump parameter, $I=|E|^2$ is the intensity distribution, and $I_s$ the intensity saturation parameter. To account for the effects of the lenses, we used a two dimensional Fourier transform to propagate the field from the image plane to the focal plane.

To determine the expected speckle contrast for a specific pinhole diameter, we ran the simulation repeatedly, using the same parameters but with different random initial conditions, until each converged to a steady state (SS) solution. For each SS solution, we calculated the far field intensity speckle pattern by multiplying it with the same random phase diffuser matrix. We then summed the intensity speckle patterns and calculated their composite speckle contrast. We also determined the laser output energy as a function of the pinhole diameter.

The results are presented in Fig 4. The results for the speckle contrast as a function of the number of simulation realizations (i.e. the number of added speckle patterns) are presented in Fig. 4(A). The different curves correspond to different pinhole diameters. As evident, for a pinhole diameter corresponding to the diffraction limit of the finite gain aperture (solid blue curve), the SS solution is a single mode and thus independent of the initial conditions. Consequently, the same intensity speckle pattern is added again and again so speckle contrast is invariant to the number of realizations. For an intermediate pinhole diameter equal to ~1.7 of the diffraction limit (dashed green curve), the SS solution is no longer single mode and varies with initial conditions, resulting in different intensity speckle patterns. Consequently, the speckle contrast decays as $C = 1/\sqrt{N}$ until eventually saturating at just above 0.5, thereby indicating that all of the uncorrelated laser modes have been adequately sampled and further realizations will not add any new uncorrelated speckle patterns to the ensemble. As the pinhole diameter increases, the speckle contrast decreases further and saturation occurs at lower values after a larger number of realizations until eventually it cannot be observed for 500 realizations (dash dot purple curve).

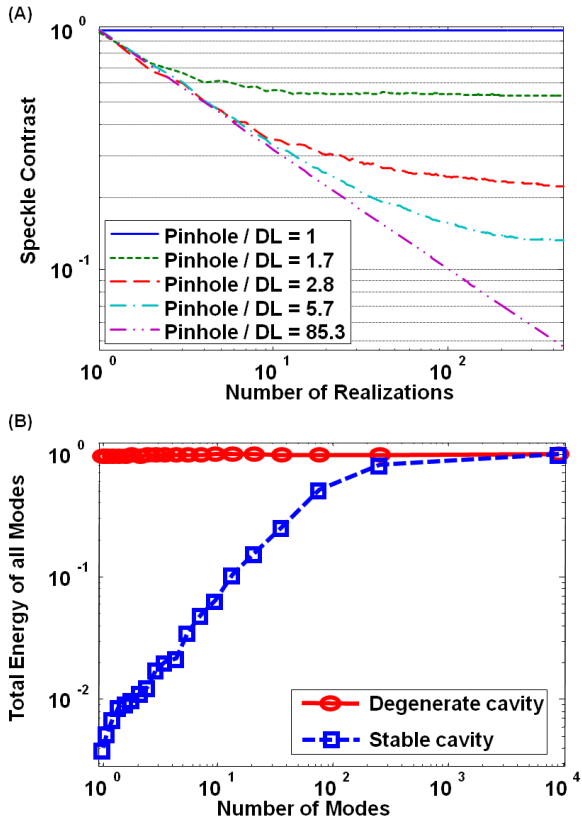

Fig.4. (Color online) Numerical simulations of speckle contrast and output energy as a function of the number of modes. (A) Calculated speckle contrast as a function of the number of simulation realizations; the different curves correspond to different pinhole diameters normalized by the diffraction limited diameter (DL). (B) Output energy as a function of the number of modes as estimated by the pinhole area over the DL area; red circles with solid line denotes the results for a degenerate cavity, blue square with dashed line denotes the results for a stable hemispherical cavity.

The simulation results of the output energy as a function of the number of modes are presented in Fig. 4(B). As evident, for the degenerate cavity (red circles with solid line), the output energy (normalized to its maximum value) varies only slightly as a function of the number of modes (calculated as the area of the pinhole over that of a diffraction limited spot). For comparison, we also include the results for a stable hemispherical internal focusing resonator (blue square with dashed line)[18]. The hemispherical internal focusing resonator[18, 19] was comprised of 2 curved mirrors ($R_1$=10 m and $R_2$=0.5 m) separated by a distance of L=46.5 cm with a spatial filtering aperture set at a distance of 2.5 cm from $R_2$. As evident, the process of mode selection within a degenerate laser cavity is exceptionally efficient, offering orders of magnitude more energy as compared to mode selection in the hemispherical internal focusing resonator.

To conclude, we demonstrated that spatial filtering in a degenerate laser cavity can be extremely efficient, enabling the laser to operate with any number of spatial modes from 1 to 320,000 with less than a factor of two variation in output energy. This unique feature can be exploited for various applications that require high brightness light sources (e.g. speckle-free full-field microscopy) or light sources with a specific spatial coherence function.

The work was supported in part by ISF-Bikura and the Weizmann Institute Yale University exchange program.